\documentclass[11pt]{article}

\usepackage{geometry}
\usepackage{currfile}
\usepackage{textcomp}
\usepackage{enumerate}

\usepackage{xcolor,soul}
\sethlcolor{yellow}

\usepackage[pdftex]{graphicx}
\DeclareGraphicsExtensions{.pdf,.jpeg,.png}
\graphicspath{{./.}}

\usepackage{fancyhdr}
\begin{document}

%\title{\Large{Revisiting the Public-Key Certificate Standard for Virtual Assets on Blockchain Systems}\\

%%\title{\Large{Towards a Public Key Management Framework for Virtual Assets\\
%% and Virtual Asset Service Providers\\

%%\title{\Large{Towards a Global Trust Network of Virtual Asset Service Providers\\

\title{\Large{Towards a Public Key Management Framework for Virtual Assets\\
and Virtual Asset Service Providers\\
~~\\
(Extended Abstract)}\\
~~}
\author{
\large{Thomas~Hardjono~~~Alexander~Lipton~~~Alex~Pentland}\\
\large{~~}\\
\large{MIT Connection Science \& Engineering}\\
\large{Massachusetts Institute of Technology}\\
\large{Cambridge, MA 02139, USA}\\
\large{~~}\\
\small{{\tt hardjono@mit.edu}~~{\tt alexlip@mit.edu}~~{\tt pentland@mit.edu}}\\
\large{~~}\\
}

\maketitle
% = = = = = = = \thispagestyle{fancy}% sets the current page style to 'fancy'

\begin{abstract}
%\boldmath
The recent FATF Recommendations defines virtual assets and virtual assets service providers (VASP),
and requires under the Travel Rule that originating VASPs obtain and 
hold required and accurate originator information and required 
beneficiary information on virtual asset transfers.
In this paper we discuss the notion of key ownership evidence
as a core part of originator and beneficiary information required by the FATF Recommendation.
We discuss approaches to securely communicate the originator and beneficiary information
between VASPs, and review existing standards for public key certificates
as applied to VASPs and virtual asset transfers.
We propose the notion of a trust network of VASPs
in which originator and beneficiary information, including key ownership information,
can be exchanged securely while observing individual privacy requirements.
~~

\end{abstract}

\newpage
\clearpage

{\small 
%\par\noindent\rule{\textwidth}{0.4pt}
\tableofcontents
%\par\noindent\rule{\textwidth}{0.4pt}
}

%\newpage
%\clearpage

~~\\

%%%%%%%%%%%%%%%%%%%%%%%%%%%%%%%%%%%%%%%%%%%%%%%%%%%%%%%%%%%%%
\section{Introduction}

Since the emergence of the Bitcoin cryptocurrency system~\cite{Bitcoin}
over a decade ago there has been a growing interest in the
use of blockchain technology as the basis for the exchange various
types of {\em virtual assets} beyond the original Bitcoin cryptocurrency~\cite{Buterin2014,SchwartzYoungs2014,TradecoinRSOS2018}.
More recently, in 2018 the Financial Action Task Force (FATF) provided a definition
of virtual assets and their service providers,
placing cryptocurrency exchanges under the category of
{\em virtual asset service providers} (VASP).
One implication, among others,
is that the existing FATF regulatory framework
applies to these exchanges,
and that exchanges must obtain and hold originator and beneficiary information
in the case of virtual asset transfers.

In this paper we review the use of existing standards in the area of public key certificates
and certificate management in the context of virtual asset service providers.
There are several goals of this paper.
The first goal of this paper is
to review the existing methods and standards
dealing with information pertaining to public keys,
key ownership and key operators.
More specifically,
we discuss the use of the existing standards for public key certificates
and the registration services used by certification authorities
as a means to obtain originator and beneficiary information prior
to the transfer of virtual assets,
thereby providing a compliant solution for these new types of service providers~\cite{Hardjono2019b}.
We also discuss the need for virtual asset service providers
to exchange customer information using the existing standards
for attributes or claims.
The public key certificates of customers,
as well as their attribute information should be communicated out-of-band (off-chain)
between virtual asset service providers.
These topics will be discussed in Section~2 to Section~7.

The second goal of this paper is to propose and discuss 
the notion of a {\em trust network} of VASPs as a way for the community of VASPs
to exchange out-of-band relevant information regarding their customers and related keys (Section~\ref{sec:Inter-VASP-Sharing}).
The trust network should be based on a common {\em operating rules}
which governs the daily running of the trust network.

Our third goal is to propose a number of areas of innovation for the nascent
virtual assets industry (Section~\ref{sec:Areas-Innovation}).
This includes new mechanisms to expand the discoverability and reachability of VASPs globally.
This will allow better connectivity and information sharing among the various VASPs around the world
-- in much the same way that ISPs in the Internet share route and endpoint reachability information
among the community of ISPs globally.

%%%%%%%%%%%%%%%%%%%%%%%%%%%%%%%%%%%%%%%%%%%%%%%%%%%%%%%%%%%%%
\section{Virtual Assets and VASPs}
\label{sec:VirtualAssetsVASPs}

The {\em Financial Action Task Force} (FATF) is an inter-governmental body 
established in 1989 by the ministers of its member countries or jurisdictions~\cite{FATF-website}.  
The objectives of the FATF are to set standards and promote effective 
implementation of legal, regulatory and operational measures 
for combating money laundering, terrorist financing and other 
related threats to the integrity of the international financial system.  
The FATF is a ``policy-making body'' which works to generate 
the necessary political will to bring about national legislative and regulatory reforms in these areas.

The FATF has developed a series of {\em Recommendations} that are recognized 
as the international standard for combating of money laundering and 
the financing of terrorism and proliferation of weapons of mass destruction.  
They form the basis for a coordinated response to the threats to the integrity 
of the financial system and help ensure a level playing field.  
First issued in 1990, the FATF Recommendations were revised in 
1996, 2001, 2003, 2012 and most recently in 2018 to ensure that 
they remain up to date and relevant, and they are intended to be of universal application.

With the emergence of blockchain technologies, virtual assets and cryptocurrencies,
the FATF recognized the need to adequately mitigate the money laundering
(ML) and terrorist financing (TF) risks associated with virtual asset activities.

In its most recent Recommendation~15~\cite{FATF-Recommendation15-2018},
the FATF defines the following:
\begin{itemize}
\item	{\em Virtual Asset}: A virtual asset is 
a digital representation of value that can be 
digitally traded, or transferred, and can be used for payment or investment purposes. 
Virtual assets do not include digital representations of fiat currencies, 
securities and other financial assets that are already covered elsewhere in the FATF Recommendations.

\item	{\em Virtual Asset Service Providers} (VASP): Virtual asset service provider means 
any natural or legal person who is not covered elsewhere under the Recommendations, 
and as a business conducts one or more of the following activities or 
operations for or on behalf of another natural or legal person:
(i) exchange between virtual assets and fiat currencies; 
(ii) exchange between one or more forms of virtual assets;
(iii) transfer of virtual assets;
(iv) safekeeping and/or administration of virtual assets or instruments enabling control over virtual assets; and
(v) participation in and provision of financial services related to an issuer's offer and/or sale of a virtual asset.
\end{itemize}
In this context of virtual assets, transfer means to conduct a transaction 
on behalf of another natural or legal person that moves 
a virtual asset from one virtual asset address or account to another.
Furthermore, 
to manage and mitigate the risks emerging from virtual assets, the Recommendations states that
countries should ensure that VASPs are regulated for AML/CFT purposes, 
and licensed or registered and subject to effective systems 
for monitoring and ensuring compliance with the relevant measures called for in 
the FATF Recommendations.

\begin{figure}[!t]
\centering
\includegraphics[width=1.0\textwidth, trim={0.0cm 0.0cm 0.0cm 0.0cm}, clip]{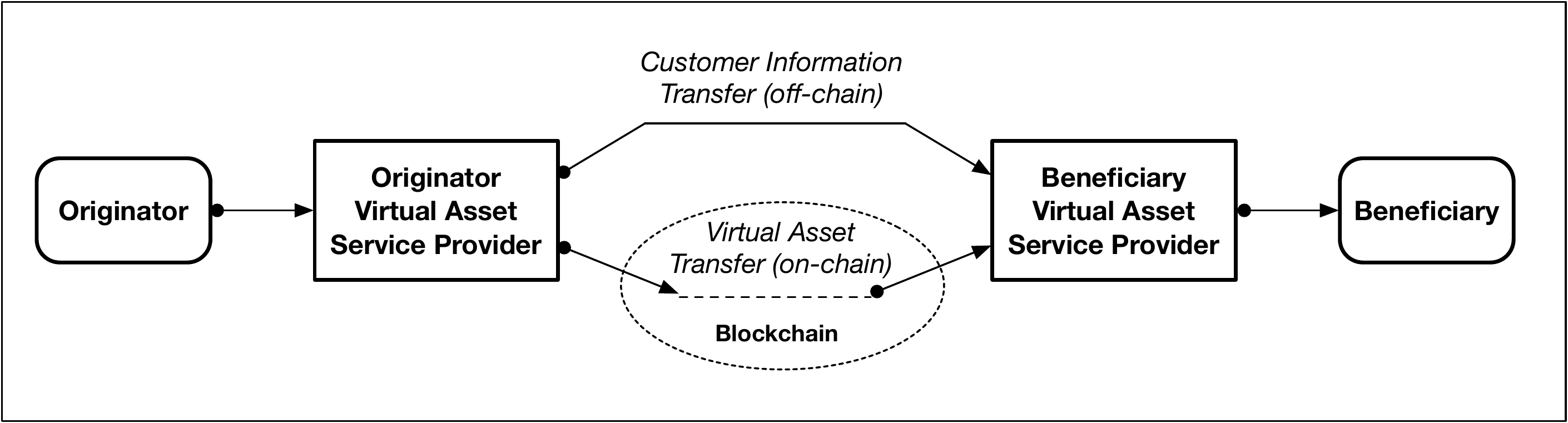}
	%
	% TRIMMING:  trim={<left> <lower> <right> <upper>} and clip options:
	% FULL EXAMPLE: \includegraphics[width=0.4\textwidth, trim={0.5cm 0.5cm 0.5cm 11.3cm}, clip]{image1.pdf}
	%
\caption{Information transfer between VASPs occurring off-chain (out-of-band)}
\label{fig:vasp2vasp-info-transfer}
\end{figure}

%%%%%%%%%%%%%%%%%%%%%%%%%%%%%%%%%%%%%%%%%%%%%%%%%%%%%%%%%%%%%
\section{The Travel Rule for Virtual Assets on Blockchains}
\label{sec:TravelRuleVASPs}

One of the key aspects of the FATF Recommendation~15
is the need for VASPs to retain information regarding the
originator and beneficiaries of virtual asset transfers:
\begin{quote}
``Countries should ensure that originating VASPs obtain and 
hold required and accurate originator information and required 
beneficiary information on virtual asset transfers, 
submit the above information to the beneficiary VASP or 
financial institution (if any) immediately and securely, 
and make it available on request to appropriate authorities. 
Countries should ensure that beneficiary VASPs obtain and hold required 
originator information and required and accurate 
beneficiary information on virtual asset transfers, 
and make it available on request to appropriate authorities. 
Other requirements of {R.16} (including monitoring of the availability of information, 
and taking freezing action and prohibiting transactions 
with designated persons and entities) apply on the same basis as set out in {R.16}. 
The same obligations apply to financial institutions 
when sending or receiving virtual asset transfers on 
behalf of a customer'' (Paragraph~7(b) of~\cite{FATF-Guidance-2019}).

\end{quote}

The implication of note~\cite{FATF-Guidance-2019} is that cryptocurrency exchanges and related VASPs
must be able to share the
originator and beneficiary information for virtual assets transactions.
This process -- also known as the {\em Travel Rule} --
originates from under the US Bank Secrecy Act (BSA - 31 USC 5311 - 5330),
which mandates that financial institutions deliver certain types of information
to the next financial institution when a funds transmittal event 
involves more that one financial institution.
This rule became effective in May 1996 and was issued by the 
Treasury Department's Financial Crimes Enforcement Network (FinCEN). 
This rule was issued by FinCEN concurrently with the new BSA record keeping rules
for funds transfers and transmittals of funds.

Given that today a virtual asset on blockchain
is controlled through the public-private keys bound to that asset,
we believe there are other information (in addition to the customer and account information)
that a VASP needs to retain in order to satisfy the travel rule:
\begin{itemize}

\item	{\em Key ownership information}: This is information pertaining to the 
legal ownership of cryptographic public-private keys.

When a customer (e.g. originator) presents their public key
to the VASP for the first time, 
there must be a ``chain of provenance'' evidence 
regarding the customer's public-private keys
which assures that the customer is the true owner.
The point is that just because an entity can prove possession of the private key,
it does not necessarily follow that the entity is 
the legal owner of the public-private keys.
Proof of possession alone is insufficient to prove legal ownership.
The ability to prove legal ownership of the public-private keys
may be crucial in different types of applications of virtual assets (e.g. property ownership).

\item	{\em Key operator information}: 
This is information or evidence pertaining to the legal custody by a VASP
of a customer's public-private keys.

This information is relevant for a VASP who
adopts a key-custody business model in which
the VASP holds and operates the customer's public-private keys
to perform transaction on behalf of the customer.

\end{itemize}

Recently there has been several objections 
on the part of some VASPs  to the FATF Recommendation~15. 
One argument is that Recommendation~15 ...``presupposes
that an originating VASP has access to other information on the
beneficiary apart from the wallet address, which it does not''~\cite{GDF2019}.
That is, some VASPs believe it is difficult or impossible for a VASP to obtain
information about the owner of a public key,
notably when the public key is the recipient (beneficiary) of an asset transfer
on a blockchain.

Another argument is that the travel rule requirement ...``can easily be circumvented by
interposing peer-to-peer (P2P) transfers or non-custodial wallets, 
which cannot be stopped''~\cite{GDF2019}.
That is, some VASPs believe that if they are burdened by the travel rule and
are forced to require wallet-holders to reveal their identity information to the VASP
(e.g. name, address, and so on, as normally required for bank accounts), 
this will force wallet-holders who wish to remain anonymous to
simply use the P2P direct asset transfer outside the regulatory framework of the FATF.

We believe these arguments are weak at best because of several reasons.
Firstly, key-holders who seek to operate outside the the FATF regulatory framework 
will continue to do so in any case.
They will continue to use the P2P direct asset transfer
(e.g. transmitting directly from their wallet to the blockchain)
regardless of VASPs or the travel rule. 
Many of these tech-savvy key-holders have been using 
the P2P direct asset transfer method
since the inception of the Bitcoin cryptocurrency 
over a decade ago~\cite{Bitcoin} -- before the notion
of crypto-exchanges came to the fore.

Secondly, the nature of virtual assets and the public-private keys
bound to those assets require careful management of those keys
(e.g. key storage, backups, retrieval, etc).
Loss or theft of keys means loss of virtual assets.
To many individuals seeking to trade in virtual assets,
the management of these keys is too burdensome.
We see this as a business opportunity for VASPs and for the virtual assets ecosystem as a whole.
VASPs need to provide business value to honest holders of virtual assets
and provide them with a user-friendly way to trade in virtual assets.

Thirdly, VASPs need to engage with existing traditional data providers
using new business models
as a means to obtain validated information regarding
candidate customers.
When an entity seeks to open a new account at a VASP,
the VASP could then validate the entity information
against these traditional sources of information.
This includes traditional financial institutions (e.g. credit rating agencies),
government institutions (e.g. motor vehicle registries)
and other traditional service providers (e.g. Telco operators).

We believe that the travel rule provides 
the emerging VASP industry with opportunity today
to develop competitive innovations around the correct identification
of owners of virtual assets
and provide them with user-friendly ways
to transfers virtual assets at a global scale.
Early attempts to integrate public key certificates
to blockchain systems have been made (e.g.~\cite{Peyton2018,PWC2017})
but more research and development need to be conducted.
Rather than weaken the travel rule to satisfy the short-term needs of
a small number of VASPs, 
governments and VASP businesses should direct research and development
into future infrastructures for virtual assets,
digital identities, public key certification and the safe management of customer keys.
We discuss several possible areas for innovation
in Section~\ref{sec:Areas-Innovation}.

%%%%%%%%%%%%%%%%%%%%%%%%%%%%%%%%%%%%%%%%%%%%%%%%%%%%%%%%%%%%%
\section{VASPs as Virtual Asset Exchanges and Key Operators}
\label{sec:VirtualAssetExchanges}

In recent years,  two popular types of VASPs have emerged,
namely {\em centralized exchanges} (CEX)
and {\em decentralized exchanges} (DEX).
A centralized exchange may or may not hold a customer's private-key.
In the case that it does,
we refer to the exchange as a ``custodial exchange'' because it has legal custody
of the public and private keys.
In this case the CEX uses the customers public-private keys
under legal custody to perform (sign) transaction sent to the cryptocurrency network.

\begin{figure}[!t]
\centering
\includegraphics[width=0.8\textwidth, trim={0.0cm 0.0cm 0.0cm 0.0cm}, clip]{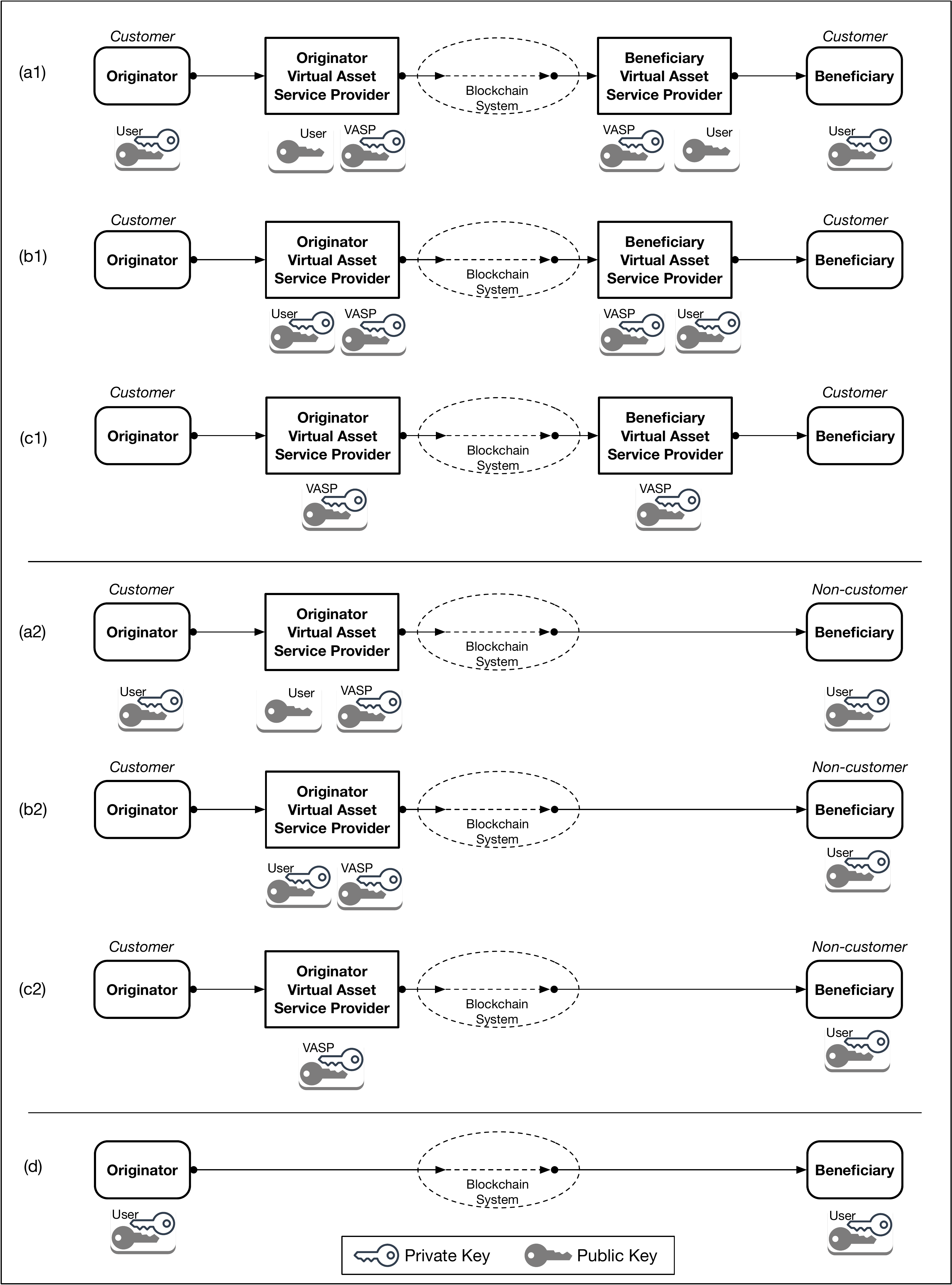}
	%
	% TRIMMING:  trim={<left> <lower> <right> <upper>} and clip options:
	% FULL EXAMPLE: \includegraphics[width=0.4\textwidth, trim={0.5cm 0.5cm 0.5cm 11.3cm}, clip]{image1.pdf}
	%
\caption{VASPs as virtual asset exchanges and key operators}
\label{fig:ExchangeTypes}
\end{figure}

In contrast, some centralized exchanges do not hold a customer's keys.
Instead, it simply creates an account for the customer in the traditional manner.
Here the CEX has one or more public-private keys of its own which it uses
to interact with the cryptocurrency network,
and it is the CEX that controls these keys (not the customer).
Furthermore, the CEX simply commingles
all their customers' funds or virtual assets
into one consolidated fund.
For the customer,
the custodial CEX approach has the attraction of relieving the 
customer from having to manage their cryptographic keys.
It also relieves the customer from having
to obtain on their own financial insurance over their virtual assets.
The CEX could obtain insurance over the entire comingled funds.

The notion of a decentralized exchange (DEX) is one where the various
functions (e.g. bid, ask, trade)
related to trading is performed the blockchain system itself
(e.g. smart contracts running on nodes of the blockchain).
The user employs a {\em wallet} (hardware and/or software)
which holds the user's public-private keys
and which performs the signing of transactions using the private key.
Here the user trades directly from their wallets without
having to trust a centralized entity.

Currently the VASP landscape is evolving and as a result there is a degree of
confusion today with regards to the notion and functions
of an exchange.
As mentioned previously in Section~\ref{sec:TravelRuleVASPs},
we believe that the travel rule will necessitate
VASPs who operate as either a centralized exchange
to address the issue of retaining
key ownership information and
key operator information.
This is particularly important for VASPs
from a risk exposure management and virtual assets 
insurance perspective~\cite{Shane2018,Dreyfuss2018}.
Figure~\ref{fig:ExchangeTypes}
attempts to summarize relationships
between VASPS, the originator/beneficiary and the location of keys
used to perform transactions.
Note that Figure~\ref{fig:ExchangeTypes}~(a1), (b1) and (c1) are symmetrical
in that there is an Originator-VASP and a Beneficiary-VASP.
In contrast,
Figure~\ref{fig:ExchangeTypes}~(a2), (b2) and (c2) are asymmetrical in that
there are no Beneficiary-VASP present.
\begin{itemize}

\item	{\em VASP mediated asset transfers}: In this model
the customer holds their public-private keys
while the VASP holds a copy of the customer's public key only (not their private key).
This model may be suitable for customers
who seek the mediation of the VASP in asset transfers (e.g. for legal purposes)
but who may not wish to provide the VASP with the customer's private key.
This is represented in Figure~\ref{fig:ExchangeTypes}~(a1) and (a2).

\item	{\em VASP key-custody asset transfers}: In this model, 
the VASPs holds custody of the customers' public-private keys.
Upon instruction from the customer,
the VASP signs transactions on behalf of the customer using the customer's private key.
This is represented in Figure~\ref{fig:ExchangeTypes}~(b1) and (b2).

\item	{\em VASP-key commingled asset transfers}:
In this model, the VASP uses its own public-private keys
to perform virtual asset transfers.
This is represented in Figure~\ref{fig:ExchangeTypes}~(c1) and (c2).

Multiple asset transfers instances could be merge into a single transaction,
thereby saving the Originator-VASP some transmission cost (e.g. gas fee).
The beneficiaries information must still be communicated
by the Originator-VASP to the Beneficiary-VASP out-of-band.

\item	{\em User direct P2P asset transfers}: In this case,
the originator and the beneficiary transacts directly on the blockchain
using their respective keys (i.e. directly from their wallets).
This is represented in Figure~\ref{fig:ExchangeTypes}~(d).
Here the user burdens the risk of key loss and thus the risk of losing their virtual assets.
VASPs are not present in this situation.

\end{itemize}
Note that combinations of models represented by
Figure~\ref{fig:ExchangeTypes}~(a1), (b1) and (c1)
can be achieved.
For example, on the sending side
the originator entity could be holding its public-private keys
as shown on the left side of Figure~\ref{fig:ExchangeTypes}~(a1),
while on the receiving side the beneficiary entity
could be using a key-custody service offered by the Beneficiary-VASP
as shown on the right side of Figure~\ref{fig:ExchangeTypes}~(b1).
Although not shown in Figure~\ref{fig:ExchangeTypes},
customers may be using other public-private keys
to establish a secure and authenticated channel between the customer
and the respective VASP.
In the remainder of this paper we will not discuss
these auxiliary keys, and focus solely on keys
that are used to transfers virtual assets on the blockchain
(i.e. the public keys that recorded on the confirmed transactions on the ledger).

%%%%%%%%%%%%%%%%%%%%%%%%%%%%%%%%%%%%%%%%%%%%%%%%%%%%%%%%%%%%%
\section{Public Key Certificates and the Travel Rule}
\label{sec:ProvingOwnership}

One of the fundamental challenges of public-key cryptography 
since its inception in 1978 ~\cite{DiffieHellman76,RivestShamirAdleman78}
is that of {\em proving ownership} of a given public key.
When two parties sign a contract or exchange signed messages,
both parties need assurance that they are employing the correct 
public keys belonging to the respective parties
(i.e. not stolen from another user).
They also need the feature of {\em non-repudiation},
meaning that a signer must be deterred or prevented from cheating
by way of claiming -- after a contract has just been signed --
that their private key was stolen before the contract was signed
(e.g. thereby repudiating the signing of the contract). 
In the context of the travel rule (Paragraph~7(b) of~\cite{FATF-Guidance-2019})
VASPs will need to retain evidence of key ownership for compliance purposes.
The existing standards pertaining to {\em public key certificates}
may provide the basis for VASPs to record information
regarding the ownership of public keys related to virtual assets.

In the late 1990s the computer industry developed the notion of
public key {\em ownership registration and certification}~\cite{HousleyPolk2001}.
The idea of registration and certification is to unambiguously 
determine the legal ownership of a public key,
and therefore provide the recipient of a signed contract or message
with a degree of {\em assurance} -- for business risk assessment --
of the true identity of the signer (the owner of the public-private keys).
The goal was to establish a {\em public key framework}~\cite{NIST-800-57,NIST-800-152}
that allowed for legal interpretation
to be created atop the framework,
where roles, responsibilities and liabilities would
be unambiguously identified and risks allocated.
The notion of a public key framework paved the way for the eventual
establishment of the e-Signature Act in year 2000~\cite{eSign2000}.

In contrast, around the same time.
some in industry sought to develop ``self-certification'' for public keys
in which the key owner
would self-declare their ownership to friends and family in a ``web of trust'' model (e.g. PGP~\cite{rfc1991}).
However, in the context of business transactions self-certification came to be viewed
as having little or no value,
and as such the ``web of trust'' philosophy failed to gain adoption in
the business community.

\begin{figure}[!t]
\centering
\includegraphics[width=0.9\textwidth, trim={0.0cm 0.0cm 0.0cm 0.0cm}, clip]{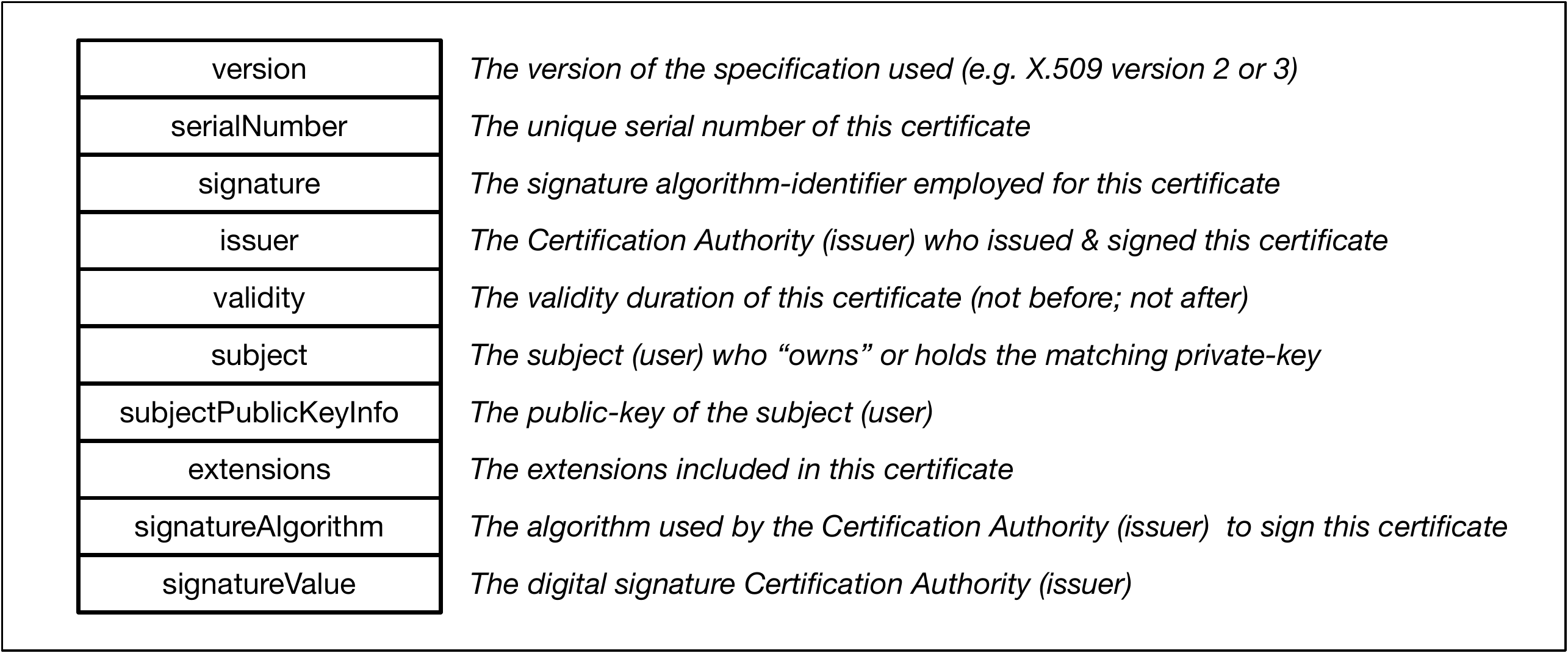}
	%
	% TRIMMING:  trim={<left> <lower> <right> <upper>} and clip options:
	% FULL EXAMPLE: \includegraphics[width=0.4\textwidth, trim={0.5cm 0.5cm 0.5cm 11.3cm}, clip]{image1.pdf}
	%
\caption{Summary of {X.509} (v3) certificates (after~\cite{rfc2246,rfc5246})}
\label{fig:X509basics}
\end{figure}

In order to understand why public key certificates are core to conducting business
on the Internet,
it is important to understand the notion of {\em technical trust} and {\em business trust}.
In order for two transacting parties to obtain assurance of each other's
key ownership,
a neutral {\em trusted third party} is needed to `` vouch'' for key ownerships
of the respective parties -- by way of 
performing ownership registration and certification of public keys.
This trusted third party is referred to as the {\em Certification Authority} (CA),
and the result of certification is a data structure 
referred to as {\em Public Key Certificates},
typically employing the {X.509} standard format.
A certification authority issues an {X.509} certificate by digitally signing
the certificate using its own private key.
By signing it, the certification authority legally {\em attests to the truthfulness}
of its assertion that the public key listed
in the {X.509} certificate is owned by the subject (person or organization) 
listed in the same certificate
(see Figure~\ref{fig:X509basics}).
One can therefore say that a certification authority ``binds''
a given public key to its owner (the subject).
The overall goal of a certification authority issuing (signing)
a public key certificate under a given public key framework
is to support the correct identification of the subject
and indirectly provide the basis for 
the {\em chain of provenance} of the public key~\cite{NIST-800-25}.
The standard protocols and formats related to public key certificates
is the {X.509} public key standard~\cite{rfc2459,rfc5280,ISO9594-pubkey} (ISO/IEC 9594-8).
Figure~\ref{fig:X509basics} summarizes the typical  {X.509} public key certificate.
The JSON-based format is defined in~\cite{rfc7517}.

The certification authority itself asserts the ownership of
its public key in the form of a {\em root certificate} 
using the same {X.509} certificate standard.
The {X.509} root certificate is typically self-signed by the certification authority
using its private key (matching the public key stated in the root certificate)~\cite{Kohnfelder1978}.
In order to prevent the certification authority from cheating by way of
modifying the root public-private keys,
the root certificate is typically made available to the broad community
in numerous ways.
This can be achieved, for example,
by publishing the root certificate in newspapers and bulletins,
by shipping copies inside browsers,
by inclusion in hardware, and so on.
In this way the certification authority is prevented
from repudiating or falsifying its own self-signed root certificate.

\begin{figure}[!t]
\centering
\includegraphics[width=1.0\textwidth, trim={0.0cm 0.0cm 0.0cm 0.0cm}, clip]{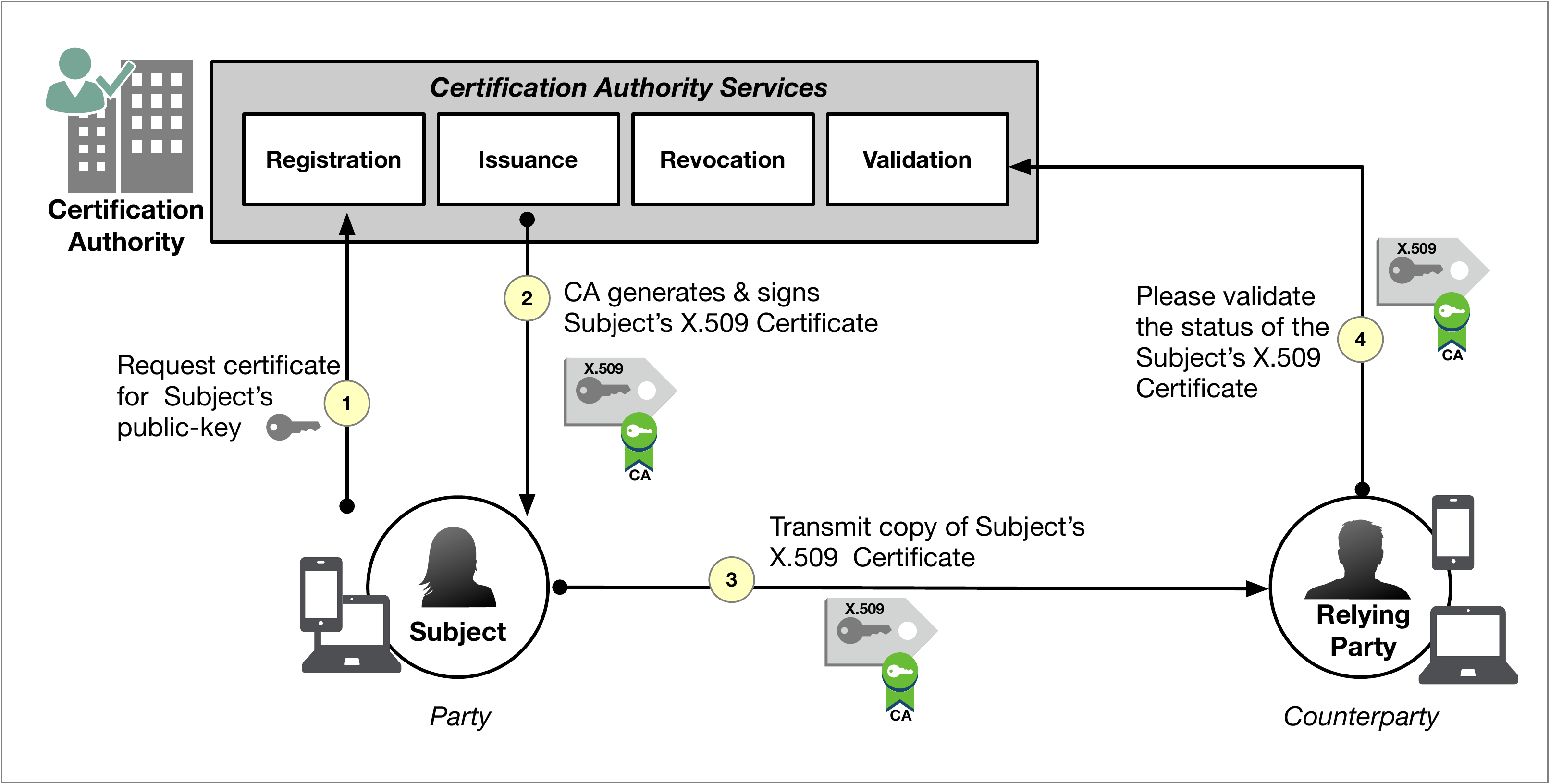}
	%
	% TRIMMING:  trim={<left> <lower> <right> <upper>} and clip options:
	% FULL EXAMPLE: \includegraphics[width=0.4\textwidth, trim={0.5cm 0.5cm 0.5cm 11.3cm}, clip]{image1.pdf}
	%
\caption{ Overview of Certification Authority (CA) services }
\label{fig:CAservices}
\end{figure}

As a neutral trusted third party, a certification authority
operates services pertaining to
the registration, certification and revocation 
of public-key ownership (Figure~\ref{fig:CAservices}).
As a legal business entity,
a commercial certification authority 
must publish (e.g. on its website) a service level agreement (SLA)
pertaining to these services.
This service agreement is referred to in industry as the 
{\em Certificate Practices Statement} (CPS)~\cite{rfc2527,rfc3647}.
Prior to registering their public key to a given certification authority,
a key owner (subject) must review the CPS document belonging to that certification authority
and determine whether the terms of the service 
in the CPS (e.g. key management procedures, liabilities, warranties for key loss, etc)
are acceptable to the key owner.
Some example of CPS statements can be found in~\cite{SymantecCPS2013} (Symantec/VeriSign)
and~\cite{AppleCPS2019} (Apple).

Over the years there has been misunderstandings 
regarding the notion of the {\em public key infrastructure} (PKI)
and regarding the expectations from such a service infrastructure.
One recurring complaint has been the ``centralized'' nature of the certification authority.
The fear is that a certification authority may have
too much control over transactions between the subject (key owner) and the relying party,
allowing it to revoke certificates
without sufficient reason, thereby rendering the key owner powerless.
We believe this main criticism misunderstands the role of the certification authority
as a neutral trusted third party with legal and financial liabilities.

More specifically,
both the subject (party) and the counter-party in a transaction must rely on the 
correct operations of the certification authority
and on its legal underpinnings as a business.
For risk management purposes,
both parties require a single entity (i.e. the certification authority) 
to take-on legal responsibility and liabilities
in the case that (i) errors exist in the public-key certificate;
(ii) in the case where the certification authority fails to revoke within reasonable time a certificate
corresponding to a stolen or lost private-key (leading to ambiguity of the status of a signed contract);
and
(iii) in the case that one or more of its services fails,
impacting loss on the part of either party (e.g. a valid certificate was revoked by mistake). 
Thus, the certification authority represents {\em centralized responsibility}.
Both parties in the transaction need this centralized responsibility in order to manage risks.
They need assurance that ``the buck stops'' at the certification authority
should certificate related errors or service failures occur.

Despite these misunderstandings the {X.509} standard for public key certificates
has been successful at a broad scale over the past two decades.
This is because in the current age of the open Internet
the public key infrastructure based on the {X.509} standard 
addresses the acute need of {\em business trust}
based on {\em legal trust} founded on the CPS contract (SLA),
which is in turn built upon the {\em technical trust}
afforded by the {X.509} key management framework.
Service contracts that carry monetary liabilities to the service operator
can only be economically viable to the operator and acceptable to customers
when the technical foundation of the services has been well understood
and has been clearly defined by sound technical standards.
Any effort to replace the {X.509} standard
will need to contend with these three aspects of trust
(business trust, legal trust and technical trust).

Today {X.509} certificates are ubiquitous across different markets, verticals and applications.
The {X.509} certificates are used extensively
within banking and finance~\cite{OpenBankingCPS2017,SWIFT-CPS-2017},
in defense and military networks~\cite{CNSS2009},
in government and federal systems~\cite{NIST-800-152,NIST-800-32},
and within many consumer electronic products
(e.g. PC computers~\cite{ApplePKI,Microsoft-PKI-website}, 
TPM hardware on laptops~\cite{HardjonoTPMEnterp2008,Microsoft-TPM-2017}, 
smartphones~\cite{ApplePKI}, USIM smartcards~\cite{Gemalto2008}, 
cable-modems~\cite{CableLabsPKI2019}, etc).
They are used extensively within routers, VPNs and other network elements.
Today in the networking industry
the Virtual Private Network (VPN) sub-segment alone 
is forecasted to reach 70 billion dollars in the next few years.
Most, if not all, websites today employ one or more
{X.509} certificates (of varying qualities) for SSL connections,
and billions of SSL connections are made every day from the user's browsers
to the various certificate-enabled websites.

\begin{figure}[!t]
\centering
\includegraphics[width=1.0\textwidth, trim={0.0cm 0.0cm 0.0cm 0.0cm}, clip]{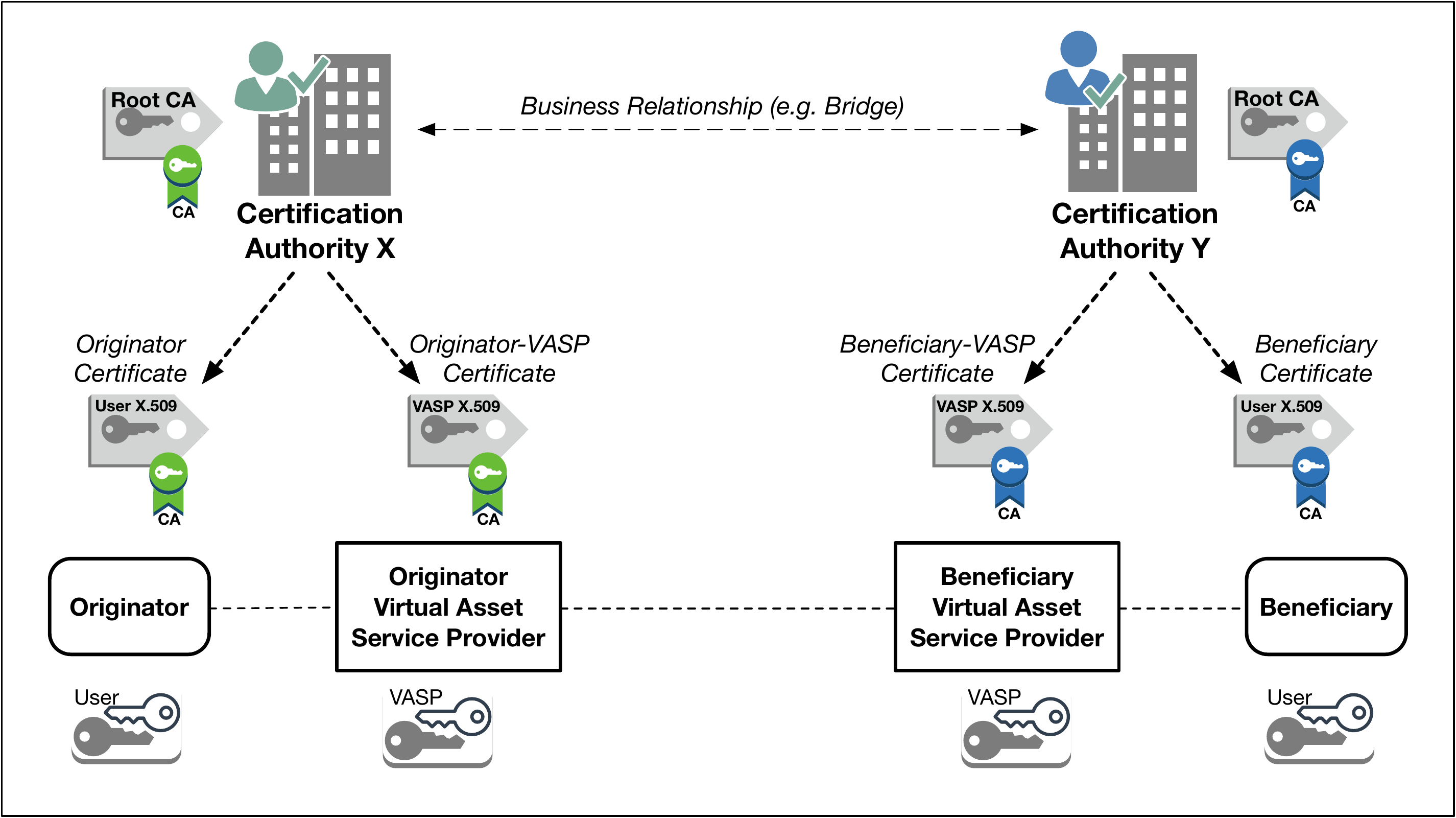}
	%
	% TRIMMING:  trim={<left> <lower> <right> <upper>} and clip options:
	% FULL EXAMPLE: \includegraphics[width=0.4\textwidth, trim={0.5cm 0.5cm 0.5cm 11.3cm}, clip]{image1.pdf}
	%
\caption{Role of certification authority for key ownership for virtual assets}
\label{fig:vasp-hierarchical}
\end{figure}

In the context of VASPs and virtual assets bound to public keys,
there are at least two approaches that VASPs can adopt with the regards
to public key certificates as a means to prove key ownership
(Figure~\ref{fig:vasp-hierarchical}):
\begin{itemize} 

\item	{\em VASP outsources customer public-key certification to a CA}: In this approach,
a VASP outsources the tasks relating to its customer's public key certificates
to an external (commercial) certification authority.
This approach allows a VASP to focus on its primary business,
leveraging the expertise of the certification authority.
All public key certificate management tasks, including certificate revocation, 
are performed by the external certification authority.

Here, when an entity (individual or organization) seeks to open an account at the VASP
accompanied by its public key
(see Figure~\ref{fig:ExchangeTypes}~(a1) and (a2)),
the VASP can redirect the entity to
the external certification authority with whom
the VASP has a business relationship.
After the entity has been successfully issued
with an {X.509} certificate for its public key,
the certification authority can provide a copy
of this certificate to the VASP.
A similar approach can be used for VASPs
which adopt a key-custody model,
in which the VASP request a public key certificate
for each customer key in its custody.

Identification information collated by the external certification authority
for the enrolling subject
can be shared with the VASP,
thereby aiding the VASP in its efforts to comply to the travel rule.

\item	{\em VASP becomes a public-key certification authority}: In this approach,
the VASP itself becomes a certification authority for its customer's public key certificates.
This approach may be attainable by a VASP depending on its business needs
and on the size of its customer base.
However, all the complex certificate management tasks must be performed by the VASP.

\end{itemize}

A third possible approach is a blend between the above,
in which a {\em hosted} certification authority 
approach is used.
Here the certificate related services are operated
by a commercial certification authority,
but the issuance of the certificates and key management
is under the full control of the VASPs.
In this case the VASP takes-on the legal liabilities
as the customer-facing certification authority.

%%%%%%%%%%%%%%%%%%%%%%%%%%%%%%%%%%%%%%%%%%%%%%%%%%%%%%%%%%%%%
\section{Off-Chain Exchange of Customer Information between VASPs}
\label{sec:attributecerts}

For virtual asset transfers,
the travel rule requires 
that originating VASPs obtain and hold 
the required and accurate originator information,
and the required  beneficiary information.
They are also required to submit this information to the beneficiary VASP 
and make it available upon request to appropriate authorities (see Paragraph~7(b) of~\cite{FATF-Guidance-2019}).
Traditionally, these include the originator's name, account number, address,
the identity of the Originator-VASP,
the amount, execution date,
and the identity of the Beneficiary-VASP.
Additionally, the side of the Beneficiary-VASP,
for incoming asset transfers the VASP must
obtain and hold
information regarding the beneficiary's name, account number, address
and other beneficiary identifiers.

In the context of blockchain systems as the medium of transaction using public keys,
the scope of information regarding the customer (originator and beneficiary)
must now include their public key information -- or what is referred to
as the ``address'' in the blockchain literature (e.g. Bitcoin~\cite{Bitcoin}).
As we suggested in Section~\ref{sec:TravelRuleVASPs},
this account information must now include
(i) key ownership information and 
(ii) key operator information
for the customer's public key used on the blockchain.
As further discussed in Section~\ref{sec:VirtualAssetExchanges}
and as illustrated in Figure~\ref{fig:ExchangeTypes},
an Originator-VASP must retain key ownership information and 
key operator information
for (a) the Originator-VASP itself,
(b) the originator customer,
(c) the Beneficiary-VASP,
and
(d) the beneficiary customer.

There are several aspects related to the collection of customer
information in the context of public keys:
\begin{itemize}

\item	{\em Customer information collected at the time certificate creation}:
Customer information must be collected prior to the issuance of their public key certificates.
Certification authorities commonly require customers (subjects)
-- whether individuals or organizations --
to submit the required information in the {\em Registration} phase
of the certificate management lifecycle~\cite{rfc2459,NIST-800-32,HousleyPolk2001}.
This is shown as Step~1 in Figure~\ref{fig:CAservices}.
During this phase it is the main task of the certification authority to perform
identity verification (of the subject) enrolling for the certificate.

\item	{\em Standardized certificate classes based on customer information provenance}:
Over the last two decade several certification authorities (CA)
have develop the notion of {\em class} or {\em grades} of public key certificates
that reflects the confidence
in the accuracy and provenance of the information regarding the customer
to whom the certificate was issued.
The classes or grades of certificates issued by a certification authority
is commonly described in the {\em Certificate Practices Statement} (CPS)~\cite{rfc2527,rfc3647}
of that certification authority.

As an emerging industry, 
VASPs can collectively define the notion of classes of certificates
for their industry
based on the required customer identification information and confidence level
during the customer registration phase.
A {\em standard definition of certificate classes for the VASP industry}
allows VASPs to require (demand) that certification authorities
fulfill the relevant customer identity verifications
and report this information and level of confidence to the VASP
as part of the service agreement.

\item	{\em VASP collation of customer identity information from the certification authority}:
In the case where a VASP outsources the issuance of certificates
to an external certification authority (CA),
the VASP must obtain a copy of the customer identity information
from the certification authority and retain it as part of  customer due diligence (CDD)
for compliance to the travel rule.

In fact, this is one reason why we believe the emerging VASP community
will benefit from closely collaborating with the long-established CA industry around the world.
Most major certification authorities
have established robust work-flows for the customer 
registration phase of the certificate issuance process.

\item	{\em Refusal of customers without certificates or uncertain identities}:
In order to comply to the requirements of the travel rule,
a VASP should simply deny customer requests for virtual asset transfers when
the customer does not posses a certificate issued by a known reliable certification authority,
or when the issuing certification authority has only low-assurance (low confidence) information regarding
the customer.

\item	{\em Certificate validation prior to virtual asset transfer}:
Prior to a virtual asset transfer, 
an Originator-VASP must perform certificate validation
of the public-key certificates of
(i) the originator customer (the customer of the Originator-VASP),
(ii) the Beneficiary-VASP,
and
(iii) the beneficiary customer (the customer of the Beneficiary-VASP).
This is discussed further in Section~\ref{sec:VerificationCertificates}.

\item	{\em Customer information communicated out-of-band between VASPs}:
As mentioned in Section~\ref{sec:VirtualAssetsVASPs}
and illustrated in Figure~\ref{fig:vasp2vasp-info-transfer},
VASPs should exchange customer information and customer certificates
(and their own VASP certificates)
out-of-band over a secure and authenticated channel.
Here, the VASP industry can standardize the APIs and connection-endpoint definitions
to allow inter-VASP exchange of customer information in a fast and efficient manner
prior to the virtual asset transfer. 
This is discussed further in Section~\ref{sec:Inter-VASP-Sharing}.

\item	{\em Standardization of customer attribute information}: 
The VASP industry should define and profile
the customer (subject) data-items required under the travel rule
to be exchanged between VASPs as part of any virtual asset transfer event.
We refer to these data items as the {\em attributes}
of the originator, beneficiary and their corresponding VASPs.

There are several standards in existence to represent attribute information
of a subject (e.g. individual or organization)
and protocols which support
the delivery of these attributes securely in the open Internet of today.
Examples include the {X.509} {\em Attribute Certificate}~\cite{rfc3281},
the XML {\em Security Assertions Mark-up Language} (SAML)~\cite{SAMLcore},
the OpenID {\em Identity Token}~\cite{OIDC1.0}
and the recent JSON-based {\em Verifiable Claims}~\cite{Sporny2019}.

\end{itemize}

\begin{figure}[!t]
\centering
\includegraphics[width=1.0\textwidth, trim={0.0cm 0.0cm 0.0cm 0.0cm}, clip]{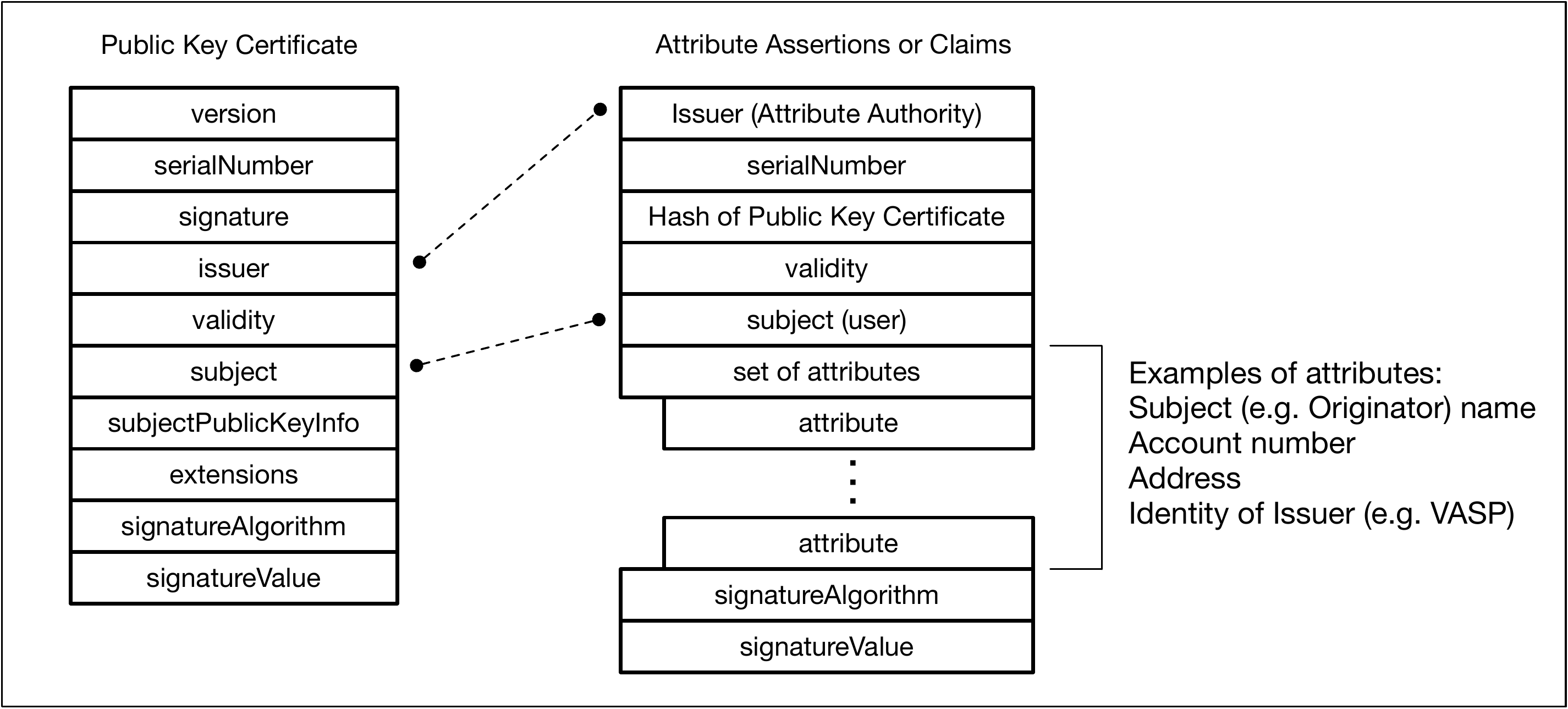}
	%
	% TRIMMING:  trim={<left> <lower> <right> <upper>} and clip options:
	% FULL EXAMPLE: \includegraphics[width=0.4\textwidth, trim={0.5cm 0.5cm 0.5cm 11.3cm}, clip]{image1.pdf}
	%
\caption{ Representation of customer information in attributes assertion }
\label{fig:CDDassertion}
\end{figure}

Following the travel rule,
there are several aspects related to the exchange of customer information and attributes between
an Originator-VASP and Beneficiary-VASP:
\begin{itemize}

\item	{\em Digitally signed by the VASP}: 
The attribute assertion or claim regarding a customer of a VASP
must be digitally signed by the VASP.
This is done as a means by the VASP to attest the truthfulness of 
the information in the assertion or claim.

\item	{\em Linking to customer public key certificate}:
A VASP that issues signed attribute assertion or claim about a customer (subject)
must link the attribute structure to the public key certificate of the customer.

This linking can be achieved by including a cryptographic hash of the customer's public key certificate
within the assertion data structure.
Figure~\ref{fig:CDDassertion} illustrates this association
using the {X.509} attribute certificates notation.

\item	{\em Preservation of subject (customer) privacy}: 
In order to safeguard customer privacy,
VASPs should limit the information exchanged to that required by the travel rule
and
to adhere to privacy regulations (e.g. GDPR~\cite{GDPR}) when maintaining customer information.
We believe there is an opportunity for the emerging VASP industry
to define and standardize the customer privacy requirements of the VASP industry
as a whole. 
We discuss this further in Section~\ref{sec:Areas-Innovation}.

When VASPs exchange attribute information out-of-band,
they must employ confidentiality measures or mechanisms
(e.g. encryption) end-to-end.
Standard protocols such as SSL or TLS~\cite{RFC2246-formatted}
which are used billions of times each day on the Internet
provides a good basis to establish
a secure and confidential channel between VASPs.

\end{itemize}

%%%%%%%%%%%%%%%%%%%%%%%%%%%%%%%%%%%%%%%%%%%%%%%%%%%%%%%%%%%%%
\section{VASP Verification of Beneficiary Public Key Certificates}
\label{sec:VerificationCertificates}

\begin{figure}[!t]
\centering
\includegraphics[width=1.0\textwidth, trim={0.0cm 0.0cm 0.0cm 0.0cm}, clip]{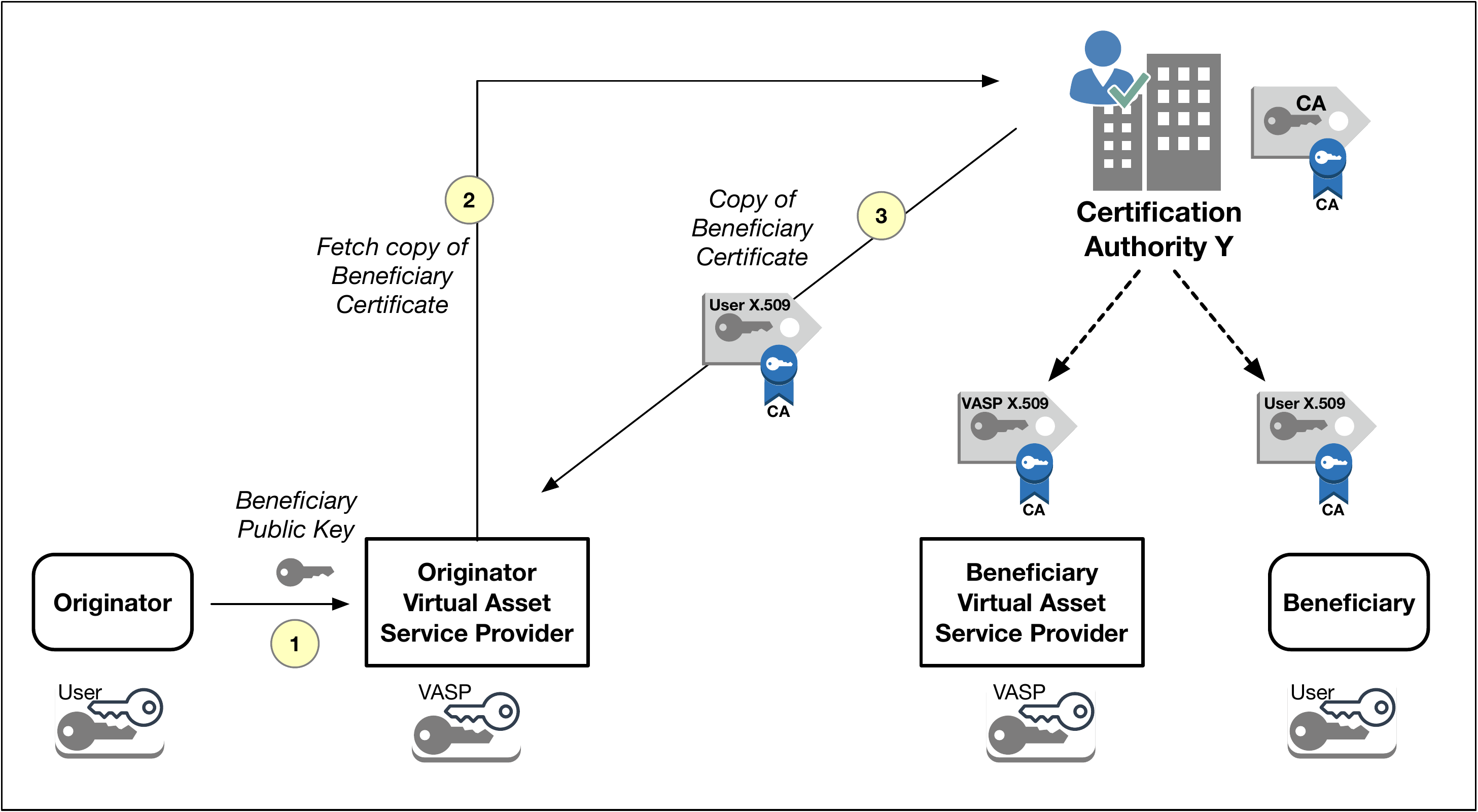}
	%
	% TRIMMING:  trim={<left> <lower> <right> <upper>} and clip options:
	% FULL EXAMPLE: \includegraphics[width=0.4\textwidth, trim={0.5cm 0.5cm 0.5cm 11.3cm}, clip]{image1.pdf}
	%
\caption{ Validating a public key certificate at its issuing certification authority (after~\cite{rfc6960}) }
\label{fig:fetchcert}
\end{figure}

As we mentioned previously,
an Originator-VASP needs to validate all relevant pubic key certificates
prior to virtual asset transfer.
This is because any errors related to the destination public keys
or about the subject identities can have dramatic impact
on the VASP from both an economic and regulatory compliance perspective.

Unlike wire transfers in correspondent banking,
transactions on a blockchain systems such as Bitcoin~\cite{Bitcoin}
is ``permanent'' (or ``immutable'' as commonly stated)
in the sense that once it is confirmed
the transaction remains in the recorded blocks (ledger) of the blockchain system.
This means that an erroneous asset transfer transaction 
cannot be
canceled, reversed or removed from the ledgers of the blockchain system
once the transaction has been confirmed.
This lack of a {\em multi-phase commitment scheme}~\cite{GrayLamport2006}
means that a mistake or error in an asset transfer to a Beneficiary-VASP
requires the Beneficiary-VASP to return
the asset in a separate transaction on the blockchain.
It is worth noting that currently most wallets and
blockchain systems generally do not employ
a phased commitment model in which
a ``pre-commit'' phase
is  followed by a ``final-commit'' phase in the sense
of classical transactional database system.
Systems such as Hyperledger Fabric~\cite{AndroulakiBarger2018} employ
Orderers and Endorsers nodes that create temporary read/write sets
prior to finalizing the read/write set of transactions.
However, this process occurs as part of the consensus-making cycle
and is outside the control of the the sender or receiver VASPs.

The implication here is that in order to avoid errors in virtual asset transfers,
in addition to verifying user account information
an Originator-VASP must validate
the origin and destination public keys of the parties {\em prior}
to broadcasting the transaction to the blockchain system.

However, there are other circumstances that may complicate this validation process:
\begin{itemize}

\item	{\em Originator possesses only its uncertified public-key}:
Many cryptocurrency users today employ a digital wallet (software and/or hardware)
that holds only the user's own public-private key and the public keys 
(``addresses'') of other users.
As such, there are cases where the originator customer may not as yet
posses a certificate for their public key.

If the originator is a customer of the VASP,
one possible course of action is for the
VASP to redirect the customer to enroll for a public key certificate
following the standard {X.509} enrollment process.
This enrollment can be done by the VASP itself if the VASP is
a certification authority.
Alternatively, the VASP can redirect its customer
to a commercial certification authority
with whom the VASP has a business relationship.

\item	{\em Originator possesses public key certificate}:
In the case that the customer provides a copy of their public key certificate,
the VASP must validate the certificate status
to the issuing certificate authority.
The {X.509} standard has protocols to perform this validation
in an efficient manner~\cite{rfc2560,rfc6960}.

\item	{\em Originator possesses only the uncertified public key of the beneficiary}:
Similar to the previous scenario,
an originator customer of a VASP may only be in possession (i.e. in their wallet)
of the public key of the beneficiary, without a corresponding certificate.

In this case, the Originator-VASP has the task of searching for
the beneficiary's certificate among other VASPs or certificate authorities.
Typically {X.509} certificates can be fetched from the issuer service
via a standardized endpoint (e.g. URI for certificates)~\cite{rfc2585,rfc4387}.

\item	{\em Originator only knows the beneficiary account information}:
In this scenario, the originator customer may only be in possession
of the beneficiary's name and account number,
and possibly the name of the destination VASP.

In this case, the Originator-VASP has the task of inquiring
to the Beneficiary-VASP about the account of the beneficiary at that VASP
using traditional means.

\end{itemize}

Figure~\ref{fig:fetchcert} illustrates the case where
an Originator-VASP obtains a copy of a certificate
based on a beneficiary public key given
by an originator customer.
The originator customer does not posses
the public key certificate of the beneficiary
but only their public key (Step~1 of Figure~\ref{fig:fetchcert}).
The Originator-VASP submits the beneficiary public key
to the relevant certification authority (Step~2),
who searches through its database of certificates.
When a match is found,
the certification authority return the certificate
of the beneficiary customer to the Originator-VASP (in Step~3).
The certification authority may also return additional
known attributes of the customer as part of the response to 
the Originator-VASP.
If the retuned certificate of the beneficiary
is valid and the returned attributes contains sufficient
information that allows the Originator-VASP to decide based on risk analysis,
the Originator-VASP may proceed with the virtual asset transfer.

In order to prevent an Originator-VASP
from querying every known certification authority in the world
-- an approach that is not only impractical but vastly inefficient --
one potential solution is for the community of VASPs
and their respective certification authorities
to form a trust network that shares known good public keys
and share certificate revocation lists.
This topic will be discussed further in
Section~\ref{sec:Inter-VASP-Sharing}.

%%%%%%%%%%%%%%%%%%%%%%%%%%%%%%%%%%%%%%%%%%%%%%%%%%%%%%%%%%%%%
\section{Towards a Trust Network of VASPs}
\label{sec:Inter-VASP-Sharing}

The Internet has been successful over the past three decades
because of a number of sound architecture designs.
One architecture design decision was to allow organizations
to own and operate portions of the Internet
as {\em autonomous systems},
allowing each autonomous system to runs its own interior routing protocol
with its own network topology for their network elements (e.g. routers, bridges, etc.).
Each autonomous system would be allocated unique identifier (i.e. AS number)
and
each autonomous system would represent independent networks (e.g. LANs, WANS, backhaul networks, etc) 
owned and operated
by various independent entities (e.g. ISPs, universities, governments, military, etc).
Thus, the Internet of today is in reality composed of numerous
autonomous systems that are ``stitched'' together, 
presenting to the user end-to-end IP connectivity.
Autonomous systems employ {\em peering agreements} or contracts
among themselves in order to negotiate IP traffic volumes and routing patterns.
These agreements permit each autonomous system
to {\em advertise routes} that are available through that autonomous system,
resulting in the reachability of (most) IP addresses globally.

Similarly,
the SWIFT banking network that begun in the 1970s
as a messaging network for sharing bank and account information
has evolved over the past three decades
into a global network that employs IP-based messaging (SwiftNet).
Instead of employing pair-wise (bilateral) key exchanges,
it has also adopted public key certificates
as a more scalable solution
for end-to-end entity authentication of members of the network~\cite{SwiftNet2004,SWIFT-CPS-2017}.

\begin{figure}[!t]
\centering
\includegraphics[width=0.8\textwidth, trim={0.0cm 0.0cm 0.0cm 0.0cm}, clip]{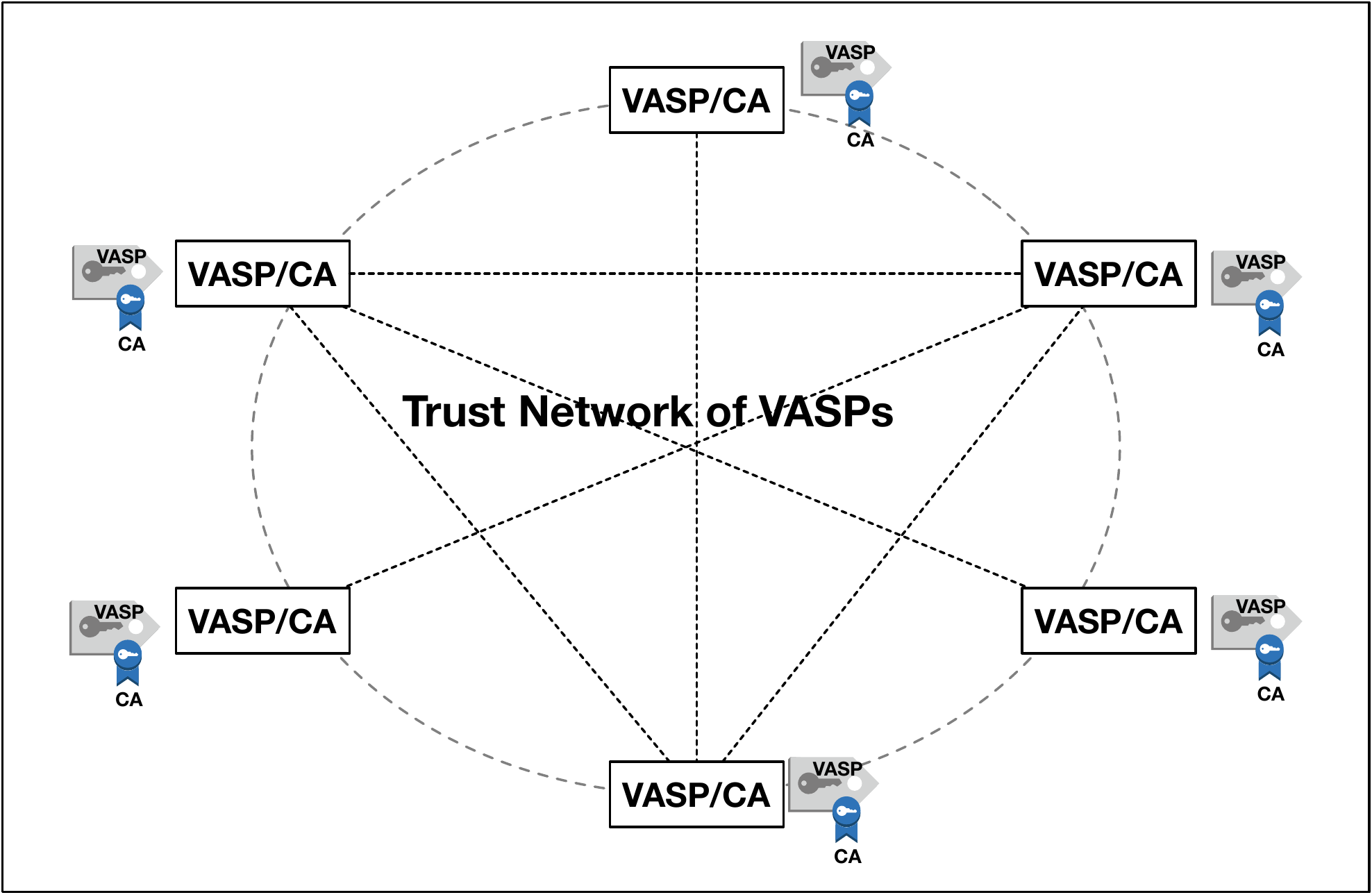}
	%
	% TRIMMING:  trim={<left> <lower> <right> <upper>} and clip options:
	% FULL EXAMPLE: \includegraphics[width=0.4\textwidth, trim={0.5cm 0.5cm 0.5cm 11.3cm}, clip]{image1.pdf}
	%
\caption{Trust Network of VASPs sharing customer certificates and attributes}
\label{fig:intra-vasp}
\end{figure}

We believe that in order to solve the various issues around the travel rule and
challenges in obtaining of originator/beneficiary customer information,
it is in the best interest of VASPs to collectively establish a {\em VASP trust network} in a manner
similar to the ISP community on the Internet (Figure~\ref{fig:intra-vasp}).

Some of the fundamental requirements of a VASP trust network are as follows:
\begin{itemize}

\item	{\em VASP-only network}:
The VASP trust network should allow the exchange
out-of-band of relevant customer-related information
as well as blockchain-transaction details.
The trust network among others
should include a technical public key framework,
a common definition of services and interfaces
and a legal framework (system rules definition)
for all its membership.

The trust network could also deploy a VASP-only blockchain system
for the purposes of common audit and reporting.
Depending on the design of this VASP-only blockchain,
hashes of the latest list of known good public keys (and pointers to their file locations)
could be captured periodically on this blockchain.

\item	{\em Independence from asset transfer blockchains}: 
The VASP trust network must be independent of any blockchain system as the medium of virtual asset transfers.
Like the Internet routing autonomous systems,
in the future there will be dozens to hundreds of blockchain systems operating around
the world (e.g. for different types of virtual assets),
presenting several challenges for blockchain interoperability~\cite{HardjonoLipton2019a}.
As an emerging industry VASPs must ensure that their trust network architecture
can interoperate with any and all future asset transfer blockchains.

\item	{\em VASP-to-VASP secure channels based on VASP public key certificates}:
In order to have the ability to quickly establish
secure and authenticated channel pair-wise between VASPs in the trust network,
the members of the trust network should each posses
public-private keys and a certificate solely for interacting on the trust network.
The use of certificates simplifies the task of communicating
the public keys of members of the trust network.
In the case that the trust network employs a VASP-only blockchain system,
then separate public-private keys must be used for that blockchain system.

Some VASPs today are already employing {X.509} certificates 
for protecting SSL connections from the customers browser to the VASP service platform.
However, this minimal use of SSL certificates needs to be enhanced
(e.g. end-to-end integration with customer wallets,
validation of chains of certificates and attribute claims,
cross-VASP certificate queries, etc.).

\end{itemize}

\begin{figure}[!t]
\centering
\includegraphics[width=1.0\textwidth, trim={0.0cm 0.0cm 0.0cm 0.0cm}, clip]{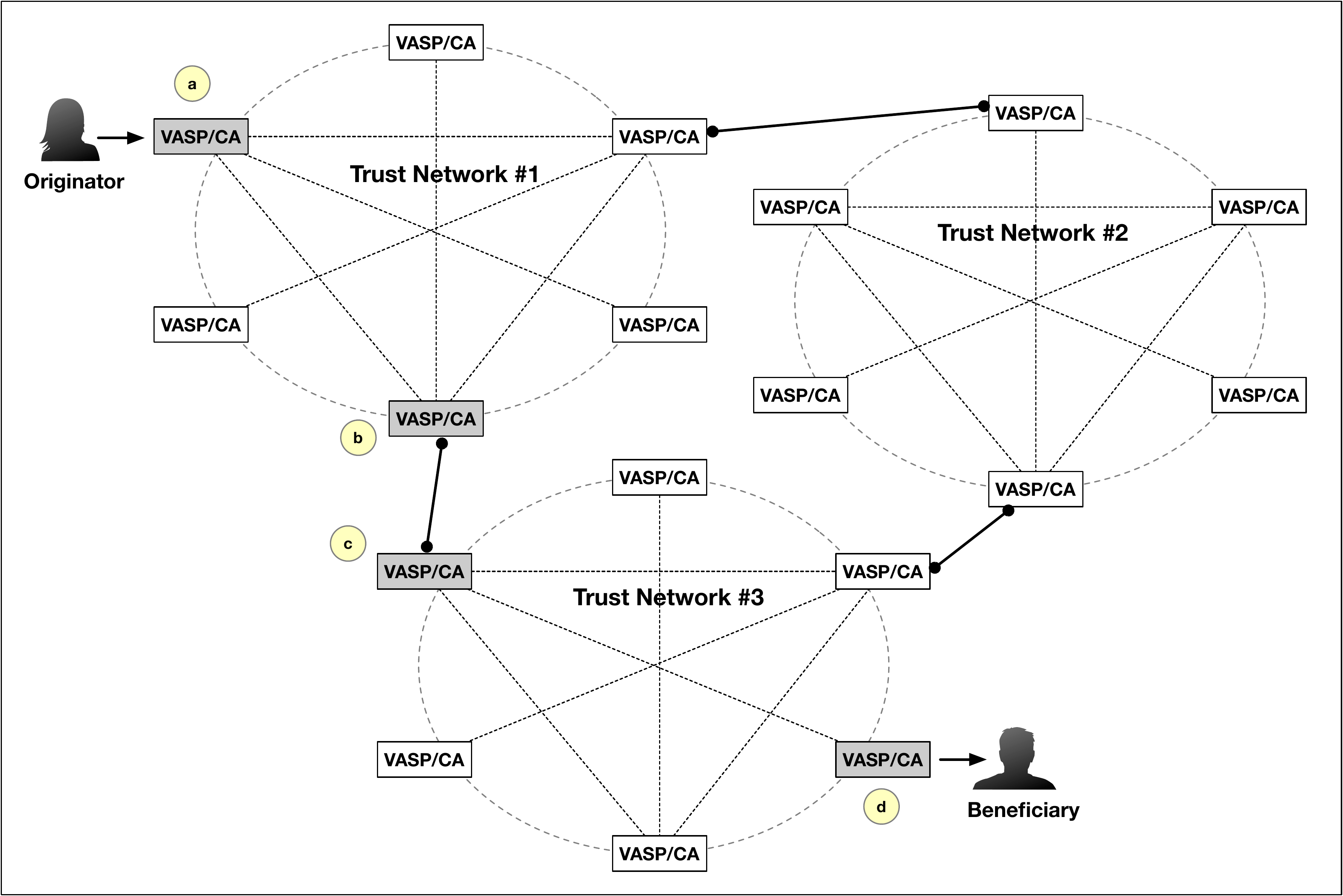}
	%
	% TRIMMING:  trim={<left> <lower> <right> <upper>} and clip options:
	% FULL EXAMPLE: \includegraphics[width=0.4\textwidth, trim={0.5cm 0.5cm 0.5cm 11.3cm}, clip]{image1.pdf}
	%
\caption{Global interconnection of multiple VASP trust networks}
\label{fig:globalintervasp}
\end{figure}

A trust network of VASPs enables and promotes virtual asset transfers globally in the following ways:
\begin{itemize}

\item	{\em Synchronization of blockchain transactions to customer identity}:
The use of a trust network running
parallel to the asset-transactions blockchain
allows an Originator-VASP to communicate to the Beneficiary-VASP ahead of the transaction on the blockchain.
The tight synchronization between the customer information sent through the trust network
and the asset transaction on the blockchain provides
the foundation for (i) post-event auditing/reconciliation, and 
(ii) evidence for conflict-resolution and among VASPs who are members of the rust network.

For example, for commingled virtual asset transfers intended for multiple individual beneficiaries,
the Originator-VASP can transmit a list of the intended amounts and beneficiaries (recipients)
through the trust network.
This provides some time (e.g. a few seconds) for the Beneficiary-VASP
to validate that all the destined beneficiaries are known
to the Beneficiary-VASP and are not on the list of suspect accounts.

\item	{\em Exchange of information about active and revoked customer certificates}:
VASPs who are members of the trust network can
exchange with each other some minimal information
regarding the public key certificates of their respective customers.

An example of this minimal information could be
the serial numbers (of the certificates) or the public key themselves (without certificates).
This list of known good serial numbers and public keys
could be shared (broadcasted) securely with
members of the trust network on a regular basis
(e.g. hourly or overnight) based on a push/pull model (e.g. over a RESTful API~\cite{rfc4387}).
This allows one VASP to query another VASP in the trust network,
submitting only the serial numbers or the public keys 
over a point-to-point secure channel (e.g. SSL).

Thus, when an originator customers provides the Originator-VASP
with a destination beneficiary public key (which the Originator-VASP may have never seen before),
the Originator-VASP can search through its copy
of the list of known good public keys in the trust network
and query the corresponding VASP in the trust network
that advertised knowledge of the matching certificate.
Similarly, the trust network of VASPs should periodically (e.g. hourly, overnight)
exchange the certificate revocation list (CRL)~\cite{rfc3280,rfc5280}
among the members of the trust network
using the existing {X.509} standard protocols.

\item	{\em Exchange of signed assertions about customers}:
When an Originator-VASP queries another VASP in the trust network
and obtains a valid copy of the public key certificate of a customer of that VASP,
the Originator-VASP has the option to further query that VASP
for additional account information regarding the customer of that VASP.
There are several standard protocols that can be used to deliver
customer information assertions or claims 
(e.g. SAML~\cite{SAMLcore} and OIDC~\cite{OIDC1.0}
are used extensively
in various identity provider communities).

The point here is that the initial sharing of minimal information 
(e.g. serial number of certificate, or public key of a customer)
among the VASPs in the trust network
allows a VASP to focus subsequent deeper queries to a single (or few) specific VASP
for obtaining the beneficiary customer information.
This increases the efficiency of the network
and the reduces the response time experienced by an Originator-VASP.

\item	{\em Global interconnection of multiple VASP trust networks}:
In order for the VASP industry to scale-up their services
towards a global customer base,
the trust network of VASPs need to be interconnected
in the same manner that autonomous systems are interconnected together
based on ISP peering contracts.
A global interconnection of multiple VASP trust networks
allows a VASP in one trust network (domain)
to obtain ``clues'' as to the existence
of another VASP in different trust network (foreign domain)
who may be in possession of information relating to a destination customer public key.

For example, in Figure~\ref{fig:globalintervasp},
the VASP at point (a) in Trust Network~1
who is seeking to fulfill an asset transfer request
from the originator in Trust Network~1 to a beneficiary in Trust Network~3
may obtain reachability knowledge about a remote VASP at point (d) within in Trust Network~3.
The expectation is that the VASP at point (d)
may posses the public key certificate and assertions 
regarding the beneficiary (who is expected to be a customer of that remote VASP).

This reachability information can be ``advertised''
from the VASPs in Trust Network~3 into the VASPs at Trust Network~1
through the peering points at VASP~(b) and VASP~(c).
That is,
the Originator-VASP at point (a) hears about Beneficiary-VASP at point (d)  
because of the ``route advertisements'' (i.e. list of public keys known in Trust Network~3)
was shared through VASP peering points at VASPs (b) and (c).

\end{itemize}

%%%%%%%%%%%%%%%%%%%%%%%%%%%%%%%%%%%%%%%%%%%%%%%%%%%%%%%%%%%%%
\section{Areas for Innovation}
\label{sec:Areas-Innovation}

There are several areas of innovation where
VASPs as an emerging industry can take leadership
and define the next-generation infrastructure for the global virtual assets commerce.

\subsection{Operating Rules for the Trust Network of VASPs}

Following from the discussion of VASP trust networks in Section~\ref{sec:Inter-VASP-Sharing},
one area of innovation for the virtual assets and VASPs industry
is the development of a common set of {\em operating rules}
suitable for the VASP industry.

Similar to other organizations (e.g. NACHA~\cite{NACHA2019}, Visa~\cite{VISA2013}, OIX~\cite{OIX2017})
the operating rules for the VASP trust network
describe a legally enforceable set rules and agreements
that govern the day-to-day running of the
trust network as a multi-party system
established to achieve common purpose of its members.
In the case of the trust network of VASPs,
these common purposes include
the sharing of:
(i) VASP entity information,
(ii) customer information,
(iii) key ownership information,
(iv) key operator information,
and 
(v) VASP reachability parameters.
These operating rules must be founded on a common set of business requirements and technical specifications.
This in turn allows each member of the trust network to obtain assurance
that each of the other participants will follow the same set of rules,
defined for their particular role in the trust network.

A good operating rules for a VASP trust network provides its
members with the several benefits.
First,
it provides a means for the members to {\em improve risk management} 
because the operating rules will allow members to quantify and manage risks
inherent in participating within the trust network.
Secondly, the operating rules provides  
its members with {\em legal certainty and predictability}
by addressing the legal rights, responsibilities, and liabilities of the participants in the trust network.
Because the operating rules is a legal agreement,
it is legally enforceable upon all members.
Thirdly, the operating rules provides
{\em transparency} to the members of the VASP trust network by making the terms of the agreements,
technical specifications (e.g. APIs, minimal performance delivery, etc), and other member business rules
-- comprising the operating rules --
accessible to all participants.

There are several business drivers for establishing a VASP trust network.
Beyond the basic set of services that members must implement
to be part of the trust network,
each member is free to offer
product/service differentiation in the market
while complying to the operating rules.
This in turn allows a VASP to broaden market adoption by enhancing
these basic services with better features (e.g. faster response,
richer customer information set,
better privacy protection for customer information, etc).
From the cost-reduction perspective,
standardizing the technical functions across all services of members of the trust network allows
for reusability of components (e.g. share common set of APIs and software libraries) 
and more efficient compliance adherence,
thus having the overall effect of lowering costs
for all members and their respective customers.

\subsection{Certificate Profiles and CPS for VASP Trust Network}

An important part of the operating rules of 
the VASP trust network is the standardization of common technical solutions
relevant to the shared goals of the trust network.
For example,
in relation to the questions of key ownership information and key operator information,
the members should develop a common
certificate practices statement (CPS) and certificate profile (CP)
for the public key certificates within the trust network.
The operating rules should define all aspects and phases of public-key management lifecycle
for all members of the VASP trust network.

There are several technical decisions regarding the certificate features
that can be defined or expressed through the certificate profile.
For example,
the certificate profile could narrow the permitted usage
of the public-private keys to that of signatures only (not encryption).
Additional VASP-specific extension could be defined
that may limit usage of the public-private keys
to only specific blockchain systems (e.g. can be used to sign transactions only for the Bitcoin network).

\subsection{Expanding the Discoverability \& Reachability of VASPs}

Following from Section~\ref{sec:Inter-VASP-Sharing},
there are several possible areas of innovation pertaining to the exchange
of VASP-related information -- within a trust network (intra-network as shown in Figure~\ref{fig:intra-vasp})
and across trust networks (inter-network as shown in Figure~\ref{fig:globalintervasp}) --
for the purpose of expanding the {\em discoverability} and {\em reachability} of VASPs.
The ability for a VASP to broadcast a query to the trust network
in search for a public key certificate of a beneficiary
represents an innovative function that promotes scaling of VASP services.
Queries should lead to responses that indicate whether an originator/beneficiary
is associated with a VASP within the local trust network,
or with a VASP in a different trust network.

Standard protocols exist today to allow access 
to certificate stores via a HTTP/SSL connection~\cite{rfc4387}.
However, additional technical extension need to be developed
that allow VASPs in a trust network 
to exchange lists of serial numbers (or valid certificates)
as well as list of public keys.
These periodic exchanges or broadcasts in the trust network
can be based on incremental changes -- so called ``deltas'' (akin to Delta CRLs in~\cite{rfc5280}) --
in order to minimize bandwidth consumption.

A given VASP may belong to multiple trust networks,
or it may have a bilateral business agreement with another VASP in a different trust network.
These VASPs could become ``gateways'' to allow certificate-related
information to flow from one trust network into another trust network,
thereby increasing the ``reach'' of the cross-network services as a whole.

\subsection{Anonymous-Verifiable Identities and Public keys in the Trust Network}

Further research and development should be devoted to a class of cryptographic schemes
that support a capability which we refer to informally 
as {\em anonymous but verifiable identities} (public keys)
with ``selective disclosure'' features.
This capability could be made available to customers of VASPs who
are members of a VASP trust network with legally-binding operating rules.
Here the cryptographic scheme should allow
an customer to posses a single private key
bound to multiple public-keys
in such a way that the public keys are {\em unlinkable} to each other when viewed by external entities.
Thus, when these public keys are used on a blockchain system,
it should be computationally difficult (infeasible)
to deduce a mathematical connection among these public keys.
The holder of such keys can prove it is a legitimate 
member of the group (i.e. member of the trust network).
We outline such an anonymous-verifiable scheme for
blockchain systems in~\cite{HardjonoPentland2016,HardjonoSmith2016a}.

There are several variants of anonymous-verifiable cryptographic 
identity schemes (e.g.~\cite{BrickellLi2012,CamenischLysyanskaya2002,Camenisch2002}).
Generally, for the application to virtual assets and VASPs,
some desirable properties are as follows:
\begin{itemize}

\item	{\em VASP trust network as group}:
In order for one VASP in a trust network to have the capacity to recognize 
an anonymous-verifiable public key wielded by a customer
of another VASP in the trust network,
all VASPs must collaborate as a group supporting the 
anonymous-verifiable scheme of choice and share the relevant
cryptographic parameters.
Depending on the specific scheme deployed,
there might be multiple overlays of the same scheme (each with different parameters)
within the same VASP trust network.

\item	{\em Parameter issuance from VASP member (Issuer-VASP)}:
In order for a customer to participate in the anonymous-verifiable identity scheme,
the customer must be registered at one of the VASPs
who is a member of the trust network implementing the scheme.
This VASP -- referred to as the Issuer-VASP --
will provide its customers with the unique relevant cryptographic parameters
that would allow the customer to transact on the blockchain
with other customers of VASPs in the same trust network
in an anonymous but verifiable manner.

Incorporated into these cryptographic parameters
is the ability for the customer to subsequently prove in an anonymous fashion
their legitimate participation (i.e. membership) in the 
anonymous-verifiable identity scheme deployed in the VASP trust network.

\item	{\em Sharing of member-verification keys among VASPs (Verifier-VASPs)}:
The VASPs in the trust network must exchange the relevant member-verification keys
so that a customer of a VASP can anonymously prove to other VASPs in the same trust network
that the customer is a legitimate customer.
These Verifier-VASPs must be able to validate the membership
of a customer in an independent manner.

\item	{\em Customer provable membership to the trust network}:
A given customer must be able to prove to any VASP in the same trust network
that the customer is a legitimate customer (of a member VASP)
in an anonymous fashion, without needing to disclose the customer's Issuer-VASP.

\item	{\em Legally enforceable registrant disclosure for specific transactions}:
When a customer is challenged with regards to
a given transaction on the blockchain (e.g. legal warrant for a suspicious transaction),
the customer and/or their Issuer-VASP must be able to
reveal the customer's true identity (e.g. to a legal court)
for that one transaction only,
while at the same time
protecting the anonymity of the other transactions belonging to that customer.

\end{itemize}
We believe these anonymous-verifiable schemes can be a potential
middle-ground solution between the need of customers to transact anonymously on the blockchain
and the need for compliance to the travel rule.

%%%%%%%%%%%%%%%%%%%%%%%%%%%%%%%%%%%%%%%%%%%%%%%%%%%%%%%%%%%%%
\section{Conclusions}

In order for the emerging virtual assets industry
to develop and evolve services that are globally accessible,
virtual assets service providers need to work collaboratively
to create the next generation infrastructures
that are not only compliant to the existing FATF regulatory framework
but also provide innovative solutions to customers globally.

Virtual assets service providers need to develop a trust network
following the principles of the Internet architecture,
allowing the exchange of customer certificates and attributes
that provide transparency into the movement of virtual assets.
The use of existing standards for public key certificates
provides a starting point for this trust network.
These standards can be extended to incorporate features
that are specific to virtual assets and to customers
of the service providers.
The overall goal is to enable originators and beneficiaries
around the world to exchange virtual assets in 
user-friendly and seamless manner,
compliant to regulations pertaining to combating
money laundering and the financing of terrorism and proliferation.

As part of developing the next generation infrastructure
the virtual assets industry should invest in research and development
in several areas of innovation.
These areas of innovation include
the development of the operating rules of the VASP trust network,
information sharing within the trust network and across trust networks,
and development of new cryptographic schemes that solve the
need of customer anonymity while complying to the requirements of the travel rule.

%------\bibliographystyle{IEEEtran}
%------\bibliography{IEEEabrv,thomasrfcbib,hardjonobib}


\end{document}